\documentclass[%
 reprint,
superscriptaddress,
nobibnotes,
 amsmath,amssymb,
 aps,
 prl
]{revtex4-2}

\usepackage{amsmath}
\usepackage{amssymb}
\usepackage{amsthm}
\usepackage{physics}
\usepackage{graphicx}
\usepackage{dcolumn}
\usepackage{bm}
\usepackage{hyperref}

\usepackage{caption}
\usepackage{subcaption}
\usepackage{mathdots}
\usepackage{mathtools}

\theoremstyle{definition}
\newtheorem{thm}{Theorem}

\newtheorem{prop}[thm]{Proposition}

\usepackage{xcolor}

\begin{document}

\preprint{APS/123-QED}

\title{Device-independent conference key agreement from two-body correlators with balanced Dicke states}

\author{Makoto Ishihara}
 \email{llmakomako.arg1076@keio.jp}
 \affiliation{%
 Department of Electronics and Electrical Engineering, Keio University, 3-14-1 Hiyoshi, Kohoku-ku, Yokohama 223-8522, Japan
}%
\author{Masahiro Takeoka}%
 \email{takeoka@elec.keio.ac.jp}
\affiliation{%
 Department of Electronics and Electrical Engineering, Keio University, 3-14-1 Hiyoshi, Kohoku-ku, Yokohama 223-8522, Japan
}%
\affiliation{
National Institute of Information and Communications Technology (NICT), Koganei, Tokyo 184-8795, Japan.
}

\date{\today}

\begin{abstract}
Device-independent conference key agreement (DI-CKA) enables $N > 2$ distant parties to establish a common secret key without detailed characterization of the source or measurement devices. Many existing protocols rely on full-body Bell correlators, whose measurement can be demanding for a large number of parties. Here, we show that two-body correlators alone suffice for DI-CKA using balanced Dicke states for every even $N \geq 4$. We construct two-body Bell inequalities and derive a tight analytical lower bound on the conditional von Neumann entropy by comparing their Bell operators with those of asymmetric Clauser-Horne-Shimony-Holt inequalities. In the ideal setting, certified key rates can approach $1-h[(N-2)/(4(N-1))]$, where $h$ is the binary entropy function. This rate remains bounded away from zero as $N \rightarrow \infty$, tending to $1-h(1/4) \approx 0.189$ bits per key-generation round. We show that the honest implementation of our protocol is invariant under collective phase noise. Moreover, a complementary full-body Bell construction certifies positive conference-key rates from correlations arbitrarily close to the local set.

\end{abstract}

\maketitle


\textit{Introduction.}---Quantum key distribution (QKD) enables two distant parties to establish a secret key with information-theoretic security~\cite{Bennett2014, Ekert1991}. Quantum conference key agreement (CKA) extends this task to $N > 2$ parties sharing a common secret key~\cite{Murta2020}. The security of standard CKA protocols relies on models of the source and the measurement devices. Deviations from these models can compromise security, as demonstrated by attacks on practical QKD systems~\cite{Lydersen2010Hacking,Gerhardt2011Full-field}. 

Device-independent CKA (DI-CKA) certifies security from a loophole-free Bell-inequality violation without requiring a detailed model of the source and the measurement devices~\cite{Ribeiro2019, Holz2020, Grasselli2023, Ishihara2025Long-distance, Wooltorton2025Genuine, Ishihara2026, Ishihara2026Enhancing}. Many existing DI-CKA protocols use Greenberger-Horne-Zeilinger (GHZ) states~\cite{Greenberger1990Bells} and Bell inequalities such as parity-Clauser-Horne-Shimony-Holt (CHSH) inequalities~\cite{Ribeiro2019} or the Holz inequality~\cite{Holz2020, Grasselli2023}. However, those Bell inequalities involve correlators of all remote parties, whose measurement can be experimentally demanding, particularly for a large number of parties.

Dicke states~\cite{Dicke1954Coherence} offer a promising alternative because their nonlocality can be detected using only few-body correlators~\cite{Tura2014Detecting, Tura2015Nonlocality}. Bell nonlocality alone, however, does not guarantee a device-independent conference key. The asymptotic DI-CKA key rate is the difference between the conditional von Neumann entropy of one remote party's (say Alice's) key bit given the side information which an eavesdropper (say Eve) may possess and the largest error-correction cost between Alice and the other remote parties. For the $X$-basis key measurements considered here, Dicke states have a nonzero error-correction cost even in the ideal setting, whereas GHZ-based protocols can have zero cost. The certified entropy must therefore be sufficiently large to compensate for these imperfect key correlations for Dicke states. Then, the following natural question arises: can we realize a DI-CKA protocol using the Dicke state and a Bell inequality involving only few-body correlators for arbitrarily many parties?

In this work, we show that two-body correlators suffice to certify a positive DI-CKA key rate using balanced Dicke states, where exactly half of the $N$ modes are excited, for every even $N \geq 4$. We construct a family of Bell inequalities and establish a tight analytical lower bound on the conditional von Neumann entropy as a function of the Bell value through an operator comparison with the asymmetric CHSH inequality developed in Ref.~\cite{Woodhead2021}. The resulting ideal key rates remain bounded away from zero as the number of parties increases. In particular, we show that an ideal DI-CKA key rate of at least $1-h(1/4) \approx 0.189$ bits per key-generation round can be certified for every even $N \geq 4$, where $h$ is the binary entropy. As a complementary result, a full-body Bell inequality certifies a positive DI-CKA key rate for correlations arbitrarily close to the local set. We also show that the honest implementation with the balanced Dicke state is invariant under collective phase noise and compare its noise tolerance with that of a GHZ-based protocol.

\textit{Protocol.}---We describe an $N$-partite DI-CKA protocol which we consider in this work. We consider $N$ parties, Alice, $\text{Bob}_1, \ldots, \text{Bob}_{N-1}$, who aim to establish a common secret key. Alice has two measurement inputs, $x \in \{0,1\}$, and each $\text{Bob}_i$ has three, $y_i \in \{0, 1, 2\}$. Let $a ~ (b_i) \in \{ \pm 1 \}$ denote Alice's ($\text{Bob}_i$'s) binary measurement outcomes. We denote the corresponding observables by $A_x$ and $B_{y_i}^i$. In the honest implementation, an $N$-qubit balanced Dicke state is distributed in each round. Note that the security analysis allows arbitrary shared states and uncharacterized devices. For even $N$, the state is
\begin{equation}\label{eq:balancedDicke}
    \ket{D_N^{N/2}} = \binom{N}{N/2}^{-1/2} \sum_{|\mathbf{z}| = N/2} \ket{\mathbf{z}},
\end{equation}
where $\mathbf{z} \in \{ 0, 1 \}^N$, $|\mathbf{z}|$ is its Hamming weight, and $\ket{\mathbf{z}}$ is the corresponding computational-basis state. Rounds with $x = 0$ and $y_i = 2$ for all $i$ are used for key generation. Rounds with $x \in \{ 0, 1 \}$ and $y_i \in \{ 0, 1\}$ are used to test the Bell inequality. Alice may apply noisy preprocessing by independently flipping each raw key bit with probability $q$ to enhance the robustness of the protocol~\cite{Ho2020}. The parties use the test rounds and a sample of the key-generation rounds to estimate an achievable key rate. Under the independent and identically distributed assumption, the asymptotic DI-CKA key rate per key-generation round $r$ is expressed as~\cite{Ribeiro2018}
\begin{equation}\label{eq:keyrate}
    r \geq  H(A|E, x = 0) - \max_i H(A|B_i, x=0, y_i = 2).
\end{equation}
Here $A$ denotes Alice's key bit after noisy preprocessing, $ B_i$ denotes $\text{Bob}_i$'s raw key bit, and $E$ denotes Eve's side information. The first term is the conditional von Neumann entropy of Alice's measurement results for the key-generation rounds given Eve's information and the second term is the maximal error-correction cost between Alice and the Bobs. If the estimated rate is positive, the parties perform error correction and privacy amplification to obtain a common secret key.

We assume that all parties use Pauli $X$ measurements for key generation in the honest implementation with the balanced Dicke state: $A_0 = B_2^1 = \cdots = B_2^{N-1} = X$. The pairwise key correlator is then $\ev{X_A X_{B_i}} = N/[2(N-1)]$. Without noisy preprocessing, the error-correction cost is
\begin{equation}\label{eq:ErrorCorrectionCost}
    H(A|B_i, x=0,y_i=2)= h(\frac{N-2}{4(N-1)}).
\end{equation}
This cost is nonzero for every even $N \geq 4$ and approaches $h(1/4) \approx 0.811$ bits as $N \rightarrow \infty$.

\textit{Two-body Bell inequalities.}---For $\alpha > 0$, define the following two-body Bell value
\begin{equation}\label{eq:TwoBodyInequality}
    \begin{split}
        S_{N, \alpha}^\text{two} = &-\sum_{i = 1}^{N-1} (\alpha \ev{A_0 (B_0^i + B_1^i)} + \ev{A_1 (B_0^i - B_1^i)}  )\\
        &- \frac{1+\alpha}{2} \sum_{1 \leq i < j \leq N-1} (\ev{B_0^i B_0^j} + \ev{B_1^i B_1^j})\\
        &- \frac{1+\alpha}{2} (N-2).
    \end{split}
\end{equation}
To relate this expression to the asymmetric CHSH inequality, let $U_y \equiv \sum_{i=1}^{N-1} B_y^i$ for $y = 0, 1$. For each fixed $y$, the observables $B_y^i$ of different Bobs commute and have eigenvalues $\pm 1$. Since $N-1$ is odd, every eigenvalue of $U_y$ is an odd integer. Thus, we can define $R_y = \text{sgn} (U_y)$, satisfying $R_y^2 = I$. Treating all the Bobs as a single composite party, we obtain the following asymmetric CHSH expression
\begin{equation}
    T_{N,\alpha}^\text{two} = - \alpha \ev{A_0 (R_0 + R_1)} - \ev{A_1 (R_0 - R_1)},
\end{equation}
with local bound $\max \{ 2, 2 \alpha\}$ and quantum maximum $2 \sqrt{1+\alpha^2}$. Let $\hat{S}$ and $\hat{T}$ denote the Bell operators associated with $S_{N, \alpha}^\text{two}$ and $T_{N,\alpha}^\text{two}$, respectively. Then, the difference between these two Bell operators can be expressed as
\begin{equation}
    \hat{T}-\hat{S} = \sum_{y=0,1} [\frac{1+\alpha}{4}(U_y^2 - I)+W_y (U_y - R_y)],
\end{equation}
where $W_0 = \alpha A_0 + A_1$, $W_1 = \alpha A_0 - A_1$, and $I$ is the identity. Let $u$ and $w$ denote the eigenvalues of $U_y$ and $W_y$, respectively. Since $[U_y, W_y] = 0$ and $||W_y|| \leq \alpha + 1$, we have
\begin{equation}
    \begin{split}
        &\frac{1+\alpha}{4} (u^2 - 1) + w(u - \text{sgn} (u))\\ \geq &\frac{1+\alpha}{4} (u^2-1) + (1+\alpha)(-|u|+1)\\
        = &\frac{1+\alpha}{4} (|u|-1)(|u|-3)\\
        \geq &0,
    \end{split}
\end{equation}
where the first inequality uses $|w| \leq \alpha + 1$ and $|u - \text{sgn}(u)| = |u|-1$, and the last inequality follows because $|u|$ is a positive odd integer. Hence $\hat{T} - \hat{S} \geq 0$, which implies
\begin{equation}\label{eq:OperatorComparison}
    S_{N, \alpha}^\text{two} \leq T_{N, \alpha}^\text{two}.
\end{equation}
We show that the local bound and the quantum maximum of $S_{N, \alpha}^\text{two}$ are $\max \{ 2, 2 \alpha \}$ and $ 2 \sqrt{1 + \alpha^2}$, respectively, in Sec. A of the Supplemental Material~\cite{SM}.

We now transfer the tight analytical entropy bound for the asymmetric CHSH inequality of Ref.~\cite{Woodhead2021} to the two-body Bell value. Define $F_\alpha (s) = \sqrt{s^2/4 - \alpha^2}$ and
\begin{equation}
\begin{split}
    g_{q, \alpha} (s) &= 1  - h(\frac{1+F_\alpha(s)}{2})\\
    &+ h(\frac{1 + \sqrt{(1-2q)^2+4q(1-q) F_\alpha(s)^2}}{2}).
\end{split}
\end{equation}
Let $g'_{q, \alpha} (s)$ denote the first derivative of $g_{q, \alpha} (s)$ and, for $0 < \alpha < 1$, let $s^*$ be the tangency point satisfying
\begin{equation}
    h(q) + g'_{q, \alpha} (s_*) (s_* -2 ) = g_{q, \alpha} (s_*).
\end{equation}
Then, the conditional von Neumann entropy is lower bounded as
\begin{equation}
    H(A|E,x=0) \geq \bar{g}_{q, \alpha} (T_{N, \alpha}^\text{two}),
\end{equation}
where
\begin{equation}\label{eq:lowerbound}
    \bar{g}_{q, \alpha} (s) = \begin{cases}
    h(q) & s \leq 2,\\
        h(q) + g'_{q, \alpha} (s_*) (s -2) & 2 < s <s_*,\\
        g_{q, \alpha} (s) &  s_* \leq s \leq 2\sqrt{1+\alpha^2},
    \end{cases}
\end{equation}
for $0 < \alpha < 1$. For $\alpha \geq 1$, the bound equals $g_{q, \alpha}(s)$ for $2 \alpha \leq s \leq 2 \sqrt{1+\alpha^2}$ and $h(q)$ for $s \leq 2 \alpha$. Since $\bar{g}_{q, \alpha}$ is nondecreasing, Eq.~(\ref{eq:OperatorComparison}) gives
\begin{equation}
    H(A|E,x=0) \geq \bar{g}_{q, \alpha} (T_{N, \alpha}^\text{two}) \geq \bar{g}_{q, \alpha} (S_{N, \alpha}^\text{two}).
\end{equation}
This bound is tight as a function of the two-body Bell value; an explicit saturating strategy is given in Sec. A of the Supplemental Material~\cite{SM}. For the ideal implementation with the balanced Dicke state specified in Sec. C of the Supplemental Material~\cite{SM}, we derive the following lower bound on the conditional von Neumann entropy for $0 < \alpha <1$, $N \geq 4$, and $q = 0$
\begin{equation}\label{eq:ExplicitLowerBound}
    H(A|E, x=0) \geq \frac{(1-\alpha)(\sqrt{1+\alpha^2}+1)}{2} - \log_2 (1+2^{-\kappa_\alpha}),
\end{equation}
where
\begin{equation}
    \kappa_\alpha = \frac{1+\sqrt{1-\alpha^2}}{\sqrt{1+\alpha^2} (\sqrt{1+\alpha^2}-1)}.
\end{equation}
As $\alpha \rightarrow 0$, the right-hand side of Eq.~(\ref{eq:ExplicitLowerBound}) approaches one. Combining this result with the error-correction cost in Eq.~(\ref{eq:ErrorCorrectionCost}), the certified ideal DI-CKA key rate approaches 
\begin{equation}\label{eq:KeyRateTwoBody}
    \lim_{\alpha \rightarrow 0}r_{N, \alpha}^\text{two} \geq 1 - h(\frac{N-2}{4(N-1)}).
\end{equation}
This approaches $1-h(1/4) \approx 0.189$ bits per key-generation round as $N \rightarrow \infty$. That is, a positive DI-CKA key rate can be obtained for every even $N \geq 4$ with the two-body Bell inequality and the balanced Dicke state in the ideal setting.

\begin{figure}[tbp]
\includegraphics[keepaspectratio, scale=0.45]{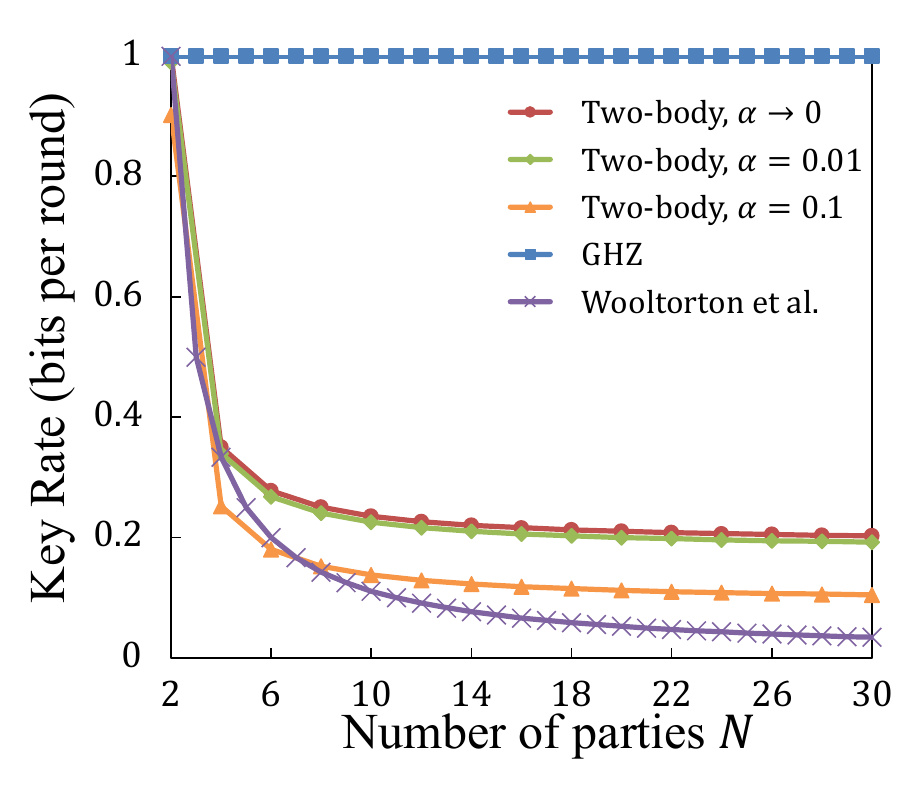}
\caption{\label{fig:VersusN}
Certified ideal key rates versus the number of parties $N$. The red circles, the green diamonds, and the orange triangles represent DI-CKA key rates of the protocol based on the balanced Dicke states with the two-body Bell inequality for $\alpha \rightarrow 0$, $\alpha = 0.01$, and $\alpha = 0.1$, respectively, where we use the explicit lower bound on the conditional von Neumann entropy in Eq.~(\ref{eq:ExplicitLowerBound}). The blue squares represent key rates of the GHZ-based protocol of Ref.~\cite{Ishihara2026Enhancing} and the purple crosses represent key rates of the $N$-partite extension of the biseparable-state DI-CKA protocol of Ref.~\cite{Wooltorton2025Genuine}. Rates are given in bits per key-generation round.}
\end{figure}

Figure~\ref{fig:VersusN} compares the certified ideal DI-CKA key rates with those of the GHZ-based protocol in Ref.~\cite{Ishihara2026Enhancing} and the biseparable-state protocol in Ref.~\cite{Wooltorton2025Genuine}. The GHZ-based protocol achieves one secret bit per round in the ideal setting, but its Bell test requires full-body correlators. The protocol of Ref.~\cite{Wooltorton2025Genuine} uses only two-body correlators and biseparable states; its straightforward $N$-party extension yields a key rate of $1/(N-1)$. In our construction, the ideal DI-CKA rate remains bounded away from zero as $N$ increases, including at fixed positive values of $\alpha$ as shown in Fig.~\ref{fig:VersusN}.

\textit{DI-CKA from arbitrarily small nonlocality}---Positive key rates from arbitrarily small nonlocality are known in bipartite device-independent QKD~\cite{Wooltorton2024Device-Independent, Farkas2024Unbounded}. We now show that this phenomenon also occurs in DI-CKA using balanced Dicke states. The relevant notion of small nonlocality is proximity of a probability distribution for measurement outcomes to the local set. Define the following set of $N$-partite local correlations $\mathcal{L}_N$
\begin{equation}
    \mathcal{L}_N = \{ P_L(a, \boldsymbol{b}|x, \boldsymbol{y}) = \sum_\lambda p(\lambda) p(a|x, \lambda) \prod_{i = 1}^{N-1}  p(b_i |y_i, \lambda)  \},
\end{equation}
where $\boldsymbol{b} = b_1, \ldots, b_{N-1}$ and $\boldsymbol{y} = y_1, \ldots, y_{N-1}$. In addition, we define the distance between two probability distributions $P$ and $Q$ as
\begin{equation}
    \mathbf{d}(P, Q) = \max_{x, \boldsymbol{y}} \frac{1}{2} \sum_{a, \boldsymbol{b}} |P(a, \boldsymbol{b}|x, \boldsymbol{y}) - Q(a, \boldsymbol{b}|x, \boldsymbol{y})|,
\end{equation}
and the distance of a probability distribution from the local set as
\begin{equation}
    \mathbf{d}(P, \mathcal{L}_N) = \inf_{P_L \in \mathcal{L}_N} \mathbf{d}(P, P_L).
\end{equation}
Then, we prove the following proposition:
\begin{prop}
    For every fixed even $N \geq 4$ and every $0<\epsilon \leq 1$, there exists a correlation $P_\epsilon$, obtainable from a balanced Dicke state, for which the following statements hold simultaneously
    \begin{equation}\label{eq:prop}
        r(P_\epsilon) \geq 1-h(\frac{N-2}{4(N-1)}), \quad \mathbf{d}(P_\epsilon, \mathcal{L}_N) \leq \frac{N-1}{2}\epsilon,
    \end{equation}
    where $r(P_\epsilon)$ represents a DI-CKA key rate for $P_\epsilon$.
\end{prop}

We sketch the proof here; details are given in Sec. E of the Supplemental Material~\cite{SM}. Consider the full-body Bell expression
\begin{equation}
    T_{N, \epsilon}^\text{full} = \ev{A_0 (C_0+C_1)} + \epsilon \ev{A_1 (C_0 - C_1)},
\end{equation}
where $C_y = \bigotimes_{i = 1}^{N-1} B_y^i$ for $y = 0, 1$ and $0 < \epsilon \leq 1$. Its local bound is 2 and its quantum maximum is $2 \sqrt{1+\epsilon^2}$. Choose $A_0 = B_2^i = X$, $A_1 = Y$, $B_0^i = M(\theta)$, and $B_1^i = M(-\theta)$, with $M(\theta) = \cos \theta X + \sin \theta Y$ and $\theta = \arctan \epsilon$. These measurements on the balanced Dicke state define $P_\epsilon$ and attain the quantum maximum, i.e., $T_{N, \epsilon}^\text{full} = 2 \sqrt{1+\epsilon^2} > 2$. Define $G_\epsilon (t) = \sqrt{t^2/4-1}/\epsilon$. For $q = 0$, the entropy bound is
\begin{equation}
    H(A|E, x=0) \geq 1 - h(\frac{1+G_\epsilon(T_{N, \epsilon}^\text{full})}{2}),
\end{equation}
We establish the tightness of this bound in Supplemental Material D~\cite{SM}. At the quantum maximum, this bound gives one bit of conditional entropy. Together with the error-correction cost in Eq.~(\ref{eq:ErrorCorrectionCost}), it certifies a key rate of $1-h[(N-2)/(4(N-1))]$ for every $0 < \epsilon \leq 1$. On the other hand, at $\epsilon = 0$, every Bob measures $X$ for all inputs. The resulting correlation $P_0$ admits an explicit local-hidden-variable model, hence $P_0 \in \mathcal{L}_N$. Moreover, for the distance between $P_\epsilon$ and $P_0$, we derive the following inequality (see Sec. E of the Supplemental Material for the derivation~\cite{SM})
\begin{equation}
    \mathbf{d} (P_\epsilon, P_0) \leq (N-1)\sin \frac{\theta}{2}.
\end{equation}
Using $\sin (\theta/2) \leq \theta/2$ and $ \theta = \arctan \epsilon \leq \epsilon$, we obtain the distance bound in Proposition 1. Thus, for each fixed even $N$, a positive DI-CKA key rate remains certified while the correlation approaches the local set.

\begin{figure}[tbp]
\includegraphics[keepaspectratio, scale=0.45]{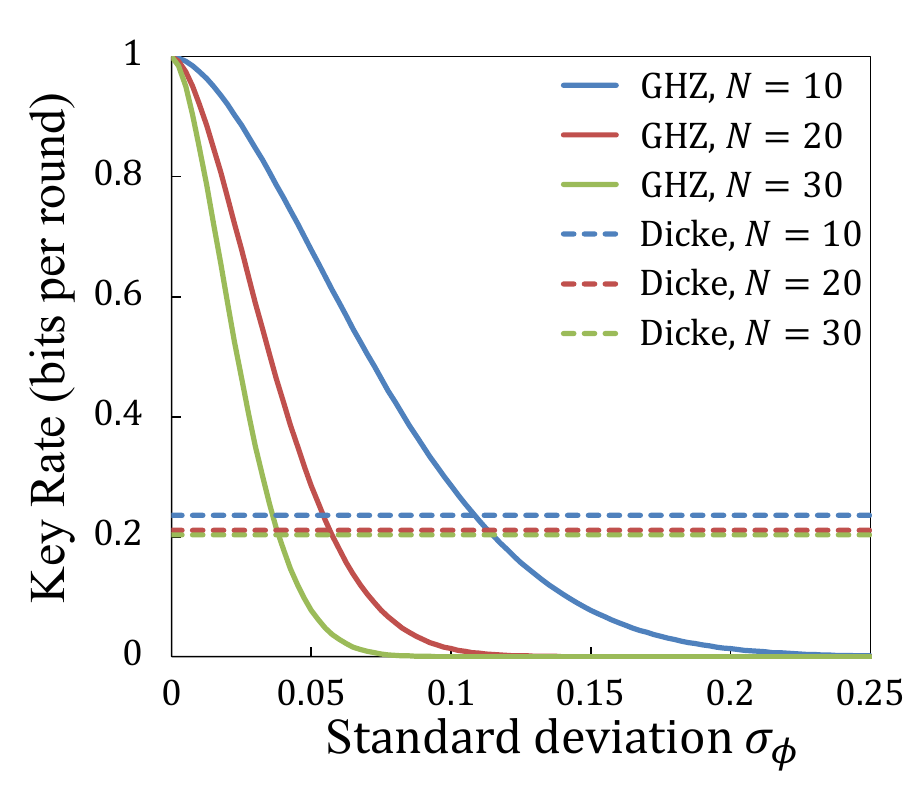}
\caption{\label{fig:VersusSigma}
Certified key rates under collective phase noise. The same random phase $\phi$ acts on all $N$ qubits, with $\phi$ drawn from a zero-mean Gaussian distribution of variance $\sigma^2_\phi$. The solid curves show the key rates for the GHZ-based protocol of Ref.~\cite{Ishihara2026Enhancing}; the dashed curves show the key rates for the two-body balanced-Dicke-based protocol in the limit $\alpha \rightarrow 0$, for $N=10, 20$ and $30$. Rates are given in bits per key-generation round.}
\end{figure}

\textit{Imperfection robustness.}---We analyze the robustness of the DI-CKA protocol based on the balanced Dicke state against experimental imperfections. We examine collective phase noise, modeled as the same random phase rotation on all $N$ qubits. For $U_\phi = e^{-i \phi Z/2}$,  an $N$-partite GHZ state, $\ket{\text{GHZ}_N} = (\ket{0}^{\otimes N} + \ket{1}^{\otimes N})/\sqrt{2}$, acquires a relative phase $N \phi$:
\begin{equation}
    U_\phi^{\otimes N} \ket{\text{GHZ}_N} = \frac{1}{\sqrt{2}} (\ket{0}^{\otimes N} + e^{i N \phi} \ket{1}^{\otimes N}),
\end{equation}
where the global phase has been omitted. Averaging over the random phase reduces the coherence of the GHZ state. In contrast, the balanced Dicke state satisfies $U_\phi^{\otimes N} \ket{D_{N}^{N/2}} = \ket{D_N^{N/2}}$. Therefore, the certified DI-CKA key rate with the honest implementation using the balanced Dicke state is invariant under this noise. Figure~\ref{fig:VersusSigma} compares the balanced-Dicke-based protocol and the GHZ-based protocol for a zero-mean Gaussian phase distribution. The key rate for the GHZ-based protocol decreases rapidly with increasing phase variance, particularly for larger $N$, whereas the key rate for the Dicke-based protocol is unchanged. We also analyze local depolarizing noise and detection inefficiency in Sec. F of the Supplemental Material~\cite{SM}. For these imperfections, the Dicke-based protocols tolerate less noise than the GHZ-based protocol. This reflects the nonzero error-correction cost for the balanced Dicke state even in the ideal setting.

\textit{Conclusion.}---In this work, we have shown that two-body Bell correlations can certify a nonvanishing DI-CKA key rate using balanced Dicke states for every even number of parties $N \geq 4$. We derive a tight analytical entropy bound by comparing the Bell operators of our two-body inequalities with those of asymmetric CHSH inequalities. The ideal key rates remain bounded away from zero as $N$ increases. A complementary full-body Bell inequality certifies a positive DI-CKA key rate for correlations arbitrarily close to the local set. In addition, the key rate with the honest implementation using the balanced Dicke state is invariant under collective phase noise, although its tolerance to local depolarizing noise and detection inefficiency is lower than that of the GHZ-based protocol considered here. An interesting future direction is to construct few-body Bell inequalities which improve tolerance to those experimental imperfections.

\textit{Acknowledgments.}---This work was supported by JST SPRING, Grant No. JPMJSP2123, JST Moonshot R\&D, Grant No. JPMJMS256E, JST CRONOS, Grant No. JPMJCS24N6, JST COI-NEXT, Grant No. JPMJPF2221 and JST ASPIRE, Grant No. JPMJAP2427. The authors used OpenAI's ChatGPT-6.1 Sol to assist with exploratory literature searches, checks of mathematical derivations, and language editing. All AI-assisted content and calculations were independently checked by the authors.

\bibliography{DIQKD}


\clearpage
\onecolumngrid

\begin{center}
\textbf{\large Supplemental Material}
\end{center}

\setcounter{equation}{0}
\renewcommand{\theequation}{S\arabic{equation}}

\section{A. Tightness of the entropy bound for the two-body Bell inequality}
We establish tightness of the entropy bound for the two-body Bell inequality by embedding a strategy that saturates the entropy bound for the asymmetric CHSH inequality. Define $\ket{\psi^-} = \frac{1}{\sqrt{2}} (\ket{01} - \ket{10})$. Let Alice, the Bobs and Eve share the following quantum state
\begin{equation}
    \rho_{A\boldsymbol{B} E} = \rho_{AB_1 E} \otimes \bigotimes_{r = 1}^{(N-2)/2} \ketbra{\psi^-}{\psi^-}_{B_{2r} B_{2r+1}},
\end{equation}
where $\rho_{AB_1E}$ corresponds to the optimal eavesdropping strategy which saturates the entropy bound for the asymmetric CHSH inequality. Here, all Bobs except $\text{Bob}_1$ measure $Z$ for both Bell-test inputs. Then, the following holds
\begin{equation}
    (B_y^{2r} + B_y^{2r + 1}) \ket{\psi^-} = 0.
\end{equation}
On the support of this state, $U_y = B_y^1$ and $R_y = B_y^1$. Hence
\begin{equation}
    \frac{1+\alpha}{4} (U_y^2 - I) + W_y(U_y - R_y) = 0,
\end{equation}
and the two Bell values coincide, i.e.,
\begin{equation}
    S_{N, \alpha}^\text{two} = T_{N, \alpha}^\text{two}.
\end{equation}
Consequently, the local bound and quantum maximum of $S_{N, \alpha}^\text{two}$ become $\max\{ 2, 2\alpha \}$ and $2 \sqrt{1+\alpha^2}$, respectively. In addition, on the support of $\rho_{A \boldsymbol{B} E}$, $T_{N, \alpha}^\text{two}$ becomes the standard asymmetric CHSH inequality, i.e.,
\begin{equation}
    T_{N, \alpha}^\text{two} = - \alpha \ev{A_0 (B_0^1 + B_1^1)} - \ev{A_1 (B_0^1 - B_1^1)}.
\end{equation}
$\rho_{AB_1E}$ saturates the entropy bound for this inequality. Thus, we can conclude that the entropy bound for the two-body Bell inequality is tight.

\section{B. An explicit entropy lower bound for $0 < \alpha < 1$}
We derive an explicit approximation to the tight entropy bound $\bar{g}_{q, \alpha}$ for $0 < \alpha < 1$. For a Bell value $s$, define
\begin{equation}
    \lambda = \frac{s-2}{2\sqrt{1+\alpha^2}-2}, \quad \kappa_\alpha = \frac{1 + \sqrt{1 - \alpha^2}}{\sqrt{1+\alpha^2} (\sqrt{1+\alpha^2}-1)},
\end{equation}
\begin{equation}
    \kappa_{q, \alpha} = \frac{1-h(q)}{(1-2q)^2} \kappa_\alpha, \quad \epsilon_{q, \alpha} = (1-2q)^2 \log_2 (1 + 2^{- \kappa_{q, \alpha}}).
\end{equation}
In the following, we derive
\begin{equation}
    \max\{h(q), h(q) + (1-h(q)) \lambda - \epsilon_{q, \alpha}\} \leq \bar{g}_{q, \alpha} (s) \leq h(q) + (1-h(q)) \lambda.
\end{equation}

Recall the analytical entropy bound
\begin{equation}
    g_{q, \alpha} (s) = 1 + h(\frac{1 + \sqrt{(1-2q)^2 + 4q(1-q) F_\alpha(s)^2}}{2}) - h(\frac{1+F_\alpha(s)}{2}),
\end{equation}
where $F_\alpha (s) = \sqrt{s^2/4 - \alpha^2}$. Define
\begin{equation}
    \Phi(z) = h(\frac{1-\sqrt{1-z}}{2}),
\end{equation}
where $z = 1 - F_\alpha(s)^2$. Then, we have
\begin{equation}
    \Phi(z) = h(\frac{1-F_\alpha(s)}{2}), \quad g_{q, \alpha}(s) = 1 + \Phi((1-(1-2q)^2)z) - \Phi(z).
\end{equation}
We first establish the concavity of $\Phi(z)$. Its first derivative is
\begin{equation}
    \Phi'(z) = \frac{1}{4u\ln 2} \ln \frac{1+u}{1-u},
\end{equation}
where $u = \sqrt{1-z}$. In addition, the second derivative becomes
\begin{equation}
    \Phi''(z) = \frac{1}{8u^3 \ln 2} (L(u) - \frac{2u}{1-u^2}),
\end{equation}
where $L(u) = \ln (1+u) - \ln (1-u)$. Now, we show $L(u) \leq \frac{2u}{1-u^2}$. We first rewrite $L(u)$ as
\begin{equation}
    \begin{split}
        \ln \frac{1+u}{1-u} &= 2 \text{arctanh}~ u\\
        &= 2 \int_0^u \frac{dt}{1 - t^2}.
    \end{split}
\end{equation}
For $0 \leq t \leq u < 1$, we have
\begin{equation}
    2 \int_0^u \frac{dt}{1-t^2} \leq 2 \int_0^u \frac{dt}{1-u^2} = \frac{2u}{1-u^2}.
\end{equation}
Then, since $8u^3 \ln 2 > 0$, we can show the concavity of $\Phi$, i.e.,
\begin{equation}
    \Phi''(z) < 0.
\end{equation}
Combining $\Phi(0) = 0$ and its concavity gives
\begin{equation}
    \begin{split}
        \Phi((1-(1-2q)^2)z) 
        &\geq (1-(1-2q)^2) \Phi(z).
    \end{split}
\end{equation}
Consequently, we have
\begin{equation}
    \begin{split}
        g_{q, \alpha}(s) &\geq 1 - (1-2q)^2 \Phi(z)\\
        &= 1 - (1- 2q)^2 h(\frac{1 - F_\alpha (s)}{2}).
    \end{split}
\end{equation}

We define the normalized distance from the quantum maximum as
\begin{equation}
    d = 1-\lambda = \frac{2\sqrt{1+\alpha^2}-s}{2\sqrt{1+\alpha^2}-2}.
\end{equation}
Using $s/2 = \sqrt{\alpha^2 + F_\alpha(s)^2}$ gives
\begin{equation}
    \begin{split}
        d &= \frac{\sqrt{1+\alpha^2} - \sqrt{\alpha^2 + F_\alpha(s)^2}}{\sqrt{1+\alpha^2} -1}\\
        &= \frac{1-F_\alpha(s)^2}{(\sqrt{1+\alpha^2} + \sqrt{\alpha^2 + F_\alpha(s)^2}) (\sqrt{1+\alpha^2}-1)}\\
        &\geq \frac{1-F_\alpha(s)^2}{2\sqrt{1+\alpha^2}(\sqrt{1+\alpha^2}-1)},
    \end{split}
\end{equation}
where we use $\sqrt{\alpha^2 + F_\alpha(s)^2}\leq \sqrt{1+\alpha^2}$ for the inequality. On the other hand, since $s \geq 2$, we have $F_\alpha(s) \geq \sqrt{1 - \alpha^2}$. Therefore,
\begin{equation}
    d \geq \frac{1 + \sqrt{1-\alpha^2}}{\sqrt{1+\alpha^2} (\sqrt{1+\alpha^2}-1)} \cdot \frac{1-F_\alpha(s)}{2} = \kappa_\alpha \cdot \frac{1-F_\alpha(s)}{2}.
\end{equation}

Define a line connecting the local bound and the quantum maximum as
\begin{equation}
    l_q(s) = h(q) + (1-h(q)) \lambda,
\end{equation}
satisfying
\begin{equation}
    \begin{split}
        l_q (s) = 1 -(1-h(q)) d.
    \end{split}
\end{equation}
We consider the difference between the lower bound $g_{q, \alpha}$ and this line as follows
\begin{equation}
    \begin{split}
        g_{q, \alpha} (s) - l_q(s) &\geq (1-h(q))d - (1-2q)^2 h(\frac{1-F_\alpha(s)}{2})\\
        &\geq (1-h(q)) \kappa_\alpha \frac{1-F_\alpha(s)}{2} - (1-2q)^2 h(\frac{1-F_\alpha(s)}{2})\\
        &= (1-2q)^2 (\kappa_{q, \alpha} \frac{1-F_\alpha(s)}{2}  - h(\frac{1-F_\alpha(s)}{2})).
    \end{split}
\end{equation}
In general, for $k \geq 0$, the first derivative of a function $f_k(e) = ke - h(e)$ is
\begin{equation}
    f_k'(e) = k - \log_2 \frac{1-e}{e}.
\end{equation}
Let $e^*$ denote a point satisfying $f'_k(e^*) = 0$. Then, we have
\begin{equation}
    \begin{split}
        e^* &= \frac{1}{1 + 2^k}.
    \end{split}
\end{equation}
Since $f_k$ is convex, $e^*$ corresponds to the minimum point. At this point, the binary entropy function becomes
\begin{equation}
    \begin{split}
        h(e^*)
        &= \log_2 (1 + 2^{-k}) + \frac{k}{1 + 2^k}.
    \end{split}
\end{equation}
Hence, the following holds
\begin{equation}
    f_k(e^*) = ke^* - h(e^*) = -\log_2 (1+2^{-k}).
\end{equation}
Consequently,
\begin{equation}
    \begin{split}
        g_{q, \alpha} (s) &\geq l_q(s) - (1-2q)^2\log_2 (1 + 2^{- \kappa_{q, \alpha}})\\
        &= l_q (s) - \epsilon_{q, \alpha}.
    \end{split}
\end{equation}

Finally, we show that this lower bound is valid for $\bar{g}_{q, \alpha}$. For $s = 2$, we have
\begin{equation}
    l_q(2) - \epsilon_{q, \alpha} = h(q) - \epsilon_{q, \alpha} \leq h(q).
\end{equation}
Therefore, $l_q(s) - \epsilon_{q, \alpha}$ lies below every convex combination of $g_{q, \alpha} (s)$ and $h(q)$. Consequently,
\begin{equation}
    \bar{g}_{q, \alpha} (s) \geq l_q(s) - \epsilon_{q, \alpha}.
\end{equation}
Furthermore, when Alice performs noisy preprocessing, we have $H(A|E, x=0) \geq h(q)$ in general. Then, we obtain
\begin{equation}
    \max\{h(q), h(q) + (1-h(q)) \lambda - \epsilon_{q, \alpha}  \} \leq \bar{g}_{q, \alpha} (s).
\end{equation}
On the other hand, using the convexity of $\bar{g}_{q, \alpha}$, $\bar{g}_{q, \alpha} (2) = h(q)$, and $\bar{g}_{q, \alpha} (2\sqrt{1+\alpha^2}) = 1$, we obtain
\begin{equation}
    \bar{g}_{q, \alpha} (s) \leq (1-\lambda) h(q) + \lambda = l_q (s).
\end{equation}

\section{C. Nonvanishing key rates for an arbitrary even number of parties}
We show that the two-body Bell inequality certifies an ideal DI-CKA key rate bounded away from zero for every even $N \geq 4$. We use the following measurements and set $q = 0$:
\begin{equation}
    A_0 = X, \quad A_1 = Z, \quad B_0^i = \cos \theta Z - \sin \theta X, \quad B_1^i = - \cos \theta Z - \sin \theta X, \quad B_2^i = X.
\end{equation}
Tracing out all but two qubits of the balanced Dicke state gives
\begin{equation}\label{appeq:twomode}
    \rho = \frac{N-2}{4(N-1)} (\ketbra{00}{00} + \ketbra{11}{11}) + \frac{N}{2(N-1)} \ketbra{D_2^1}{D_2^1}.
\end{equation}
Then, we have
\begin{equation}
    \ev{XX} = \frac{N}{2(N-1)}, \quad \ev{ZZ} = -\frac{1}{N-1}, \quad \ev{ZX} = \ev{XZ} = 0.
\end{equation}
Therefore, with the above measurements, the two-body Bell value can be expressed as
\begin{equation}
    \begin{split}
        S_{N, \alpha}^\text{two} 
        &= \alpha N \sin \theta + 2 \cos \theta - \frac{1+\alpha}{2} \frac{N^2-4}{2} \sin^2 \theta.
    \end{split}
\end{equation}
Consequently, the key rate becomes
\begin{equation}
    r_{N, \alpha}^\text{two} \geq \bar{g}_{0, \alpha} (S_{N, \alpha}^\text{two}) - h(\frac{N-2}{4(N-1)}).
\end{equation}

Choosing $\sin \theta = 2\alpha/N$, we obtain the following bound for $0 < \alpha < 1$ and even $N \geq 4$:
\begin{equation}
    \begin{split}
        S_{N, \alpha}^\text{two} &= 2 \alpha^2  + 2 \sqrt{1-\frac{4 \alpha^2}{N^2}} - \frac{(1+\alpha) (N^2 -4)}{4} \frac{4 \alpha^2}{N^2}\\
        &\geq 2 + \alpha^2 (1-\alpha) + \frac{4 \alpha^3}{N^2} -\frac{16 \alpha^4}{N^4}\\
        &= 2 + \alpha^2 (1-\alpha) + \frac{4 \alpha^3}{N^2} (1-\frac{4 \alpha}{N^2})\\
        &\geq 2 + \alpha^2 (1-\alpha),
    \end{split}
\end{equation}
where, for the first inequality, we use
\begin{equation}
    \sqrt{1 - \frac{4 \alpha^2}{N^2}} \geq 1 - \frac{2 \alpha^2}{N^2} - \frac{8\alpha^4}{N^4},
\end{equation}
and the last inequality holds for $0 < \alpha < 1$ and $N \geq 4$. Therefore, by using the explicit lower bound which we obtain in Sec. B of the Supplemental Material, we have the following lower bound on the conditional von Neumann entropy
\begin{equation}
    \begin{split}
        H(A|E, x=0) &\geq \frac{\alpha^2 (1-\alpha)}{2\sqrt{1+\alpha^2} -2} -\epsilon_{0, \alpha}\\
        &= \frac{(1-\alpha)(\sqrt{1+\alpha^2}+1)}{2} -\epsilon_{0, \alpha}.
    \end{split}
\end{equation}
Then, the DI-CKA key rate for the two-body Bell inequality becomes
\begin{equation}
    r_{N, \alpha}^\text{two} \geq \frac{(1-\alpha) (\sqrt{1+\alpha^2}+1)}{2} - \epsilon_{0, \alpha} - h(\frac{N-2}{4(N-1)}).
\end{equation}
Since $\epsilon_{0, \alpha} \rightarrow 0$ for $\alpha \rightarrow 0$, the key rate in this limit becomes
\begin{equation}
    r_{N, \alpha}^\text{two} \rightarrow 1 - h(\frac{N-2}{4(N-1)}).
\end{equation}

\section{D. Tightness of the entropy bound for the full-body Bell inequality}
In this section, we introduce a full-body Bell inequality which the balanced Dicke state maximally violates and derive a corresponding tight analytical lower bound on the conditional von Neumann entropy. 

The balanced Dicke state admits a maximally entangled decomposition between Alice's qubit and a two-dimensional logical subspace of the Bobs' joint system:
\begin{equation}
    \ket{D_N^{N/2}} = \frac{1}{\sqrt{2}} (\ket{0} \ket{D_{N-1}^{N/2}} + \ket{1} \ket{D_{N-1}^{N/2-1}}).
\end{equation}
Defining $\ket{0_L} = \ket{D_{N-1}^{N/2-1}}$ and $\ket{1_L} = \ket{D_{N-1}^{N/2}}$, this becomes $(\ket{0}\ket{1_L} + \ket{1} \ket{0_L})/\sqrt{2}$. In addition, we consider the following measurement
\begin{equation}\label{appeq:MeasurementXY}
    M(\theta) = \cos \theta X + \sin \theta Y,
\end{equation}
satisfying $M(\theta) \ket{0} = e^{i \theta}\ket{1}$ and $M(\theta) \ket{1} = e^{-i \theta}\ket{0}$. For a Dicke state of $N-1$ modes with $k$ excitations, the following holds
\begin{equation}
    M(\theta)^{ \otimes N-1} \ket{D_{N-1}^{k}} = e^{i \theta(N-1-2k)} \ket{D_{N-1}^{N-1-k}}.
\end{equation}
Thus, for the logical qubits, we have $M(\theta)^{\otimes N-1} \ket{0_L} = e^{i \theta} \ket{1_L}$ and $M(\theta)^{\otimes N-1} \ket{1_L} = e^{-i\theta}\ket{0_L}$, that is, the product of the Bobs' local outcomes realizes the corresponding logical observable. Using this gives the following for full-body correlators for the balanced Dicke state
\begin{equation}
    \ev{M_A (\gamma) \bigotimes_{i = 1}^{N-1} M_{B_i} (\beta)}_D = \cos (\gamma - \beta),
\end{equation}
where $\gamma$ is Alice's measurement angle and $\beta$ is the common measurement angle of all Bobs.

Using this logical representation, we define the following full-body Bell value
\begin{equation}\label{appeq:FullBodyInequality}
    S_{N, \alpha}^\text{full} = \alpha (\ev{A_0 C_0} + \ev{A_0 C_1}) + \ev{A_1C_0} - \ev{A_1 C_1},
\end{equation}
where $0 < \alpha$ and $C_y = \bigotimes_{i = 1}^{N-1} B_y^i$ for $y = 0, 1$. This can be understood as a variant of the asymmetric CHSH inequality developed in Ref.~\cite{Woodhead2021}. Therefore, the local bound and the quantum maximum of this Bell value are $\max \{2, 2\alpha  \}$ and $2 \sqrt{1 + \alpha^2}$, respectively. The quantum maximum can be obtained with the balanced Dicke state and the following measurements: $A_0= X$, $A_1 = Y$, $B_0^i = M(\theta)$, and $B_1^i = M(-\theta)$, where $\theta = \arctan \frac{1}{\alpha}$. Thanks to the equivalence of the above inequality and the asymmetric CHSH, we can utilize the analytical lower bound on the conditional von Neumann entropy, i.e., the following holds
\begin{equation}
    H(A|E, x=0) \geq \bar{g}_{q, \alpha} (S_{N, \alpha}^\text{full}).
\end{equation}
For the maximal quantum violation, $S_{N, \alpha}^\text{full} = 2\sqrt{1+\alpha^2}$, the lower bound becomes 1. Therefore, with the error-correction cost in Eq.~(\ref{eq:ErrorCorrectionCost}), the DI-CKA key rate in the ideal setting is
\begin{equation}\label{appeq:KeyRateFullBody}
    r_{N, \alpha}^\text{full} \geq 1 - h(\frac{N-2}{4(N-1)}),
\end{equation}
for arbitrary even $N$ and $q = 0$. 

In the following, we show the tightness of the above entropy bound for the full-body Bell inequality. We assume that Alice, the Bobs, and Eve share the following entangled state
\begin{equation}
    \ket{\Psi}_{A\boldsymbol{B}E} = \sqrt{p} \ket{D_N^{N/2}}_{A \boldsymbol{B}} \ket{0}_E + \sqrt{1-p} X_A \ket{D_N^{N/2}}_{A \boldsymbol{B}} \ket{1}_E,
\end{equation}
where $\boldsymbol{B}$ denotes the Bobs' systems $B_1, \ldots, B_{N-1}$ and $p \in [1/2, 1]$. Let the legitimate parties perform the following measurements
\begin{equation}
    A_0 = X, \quad A_1 = Y, \quad B_0^i = M(\theta_p), \quad B_1^i = M(-\theta_p), \quad B_2^i = X.
\end{equation}
From the definition of the conditional von Neumann entropy, we have
\begin{equation}
    H(A|E, x=0) = H(A|x=0) - H(E) + H(E|A,x=0).
\end{equation}
Since $\text{Tr}_{\boldsymbol{B}} [\ketbra{D_{N}^{N/2}}{D_N^{N/2}}] = \frac{1}{2}I$ and $X^2 = I$ for the identity matrix $I$, we have
\begin{equation}
    \rho_A = \text{Tr}_{\boldsymbol{B} E} [\ketbra{\Psi}{\Psi}] = \frac{1}{2}I.
\end{equation}
That is, $H(A|x=0) = 1$. On the other hand, a quantum state on Eve's subsystem can be expressed as
\begin{equation}
   \rho_E = \text{Tr}_{A \boldsymbol{B}} [\ketbra{\Psi}{\Psi}] = p\ketbra{0}{0}_E + (1-p) \ketbra{1}{1}_E. 
\end{equation}
Therefore, we have $H(E) = h(p)$.

We consider a classical-quantum state between Alice's measurement outcome and Eve's system. Tracing out the Bobs' systems from $\ket{\Psi}$ gives
\begin{equation}
    \begin{split}
        \text{Tr}_{\boldsymbol{B}} [\ketbra{\Psi}{\Psi}] 
        &= \frac{1}{2} (p I_A \ketbra{0}{0}_E + \sqrt{p (1-p)} X_A \ketbra{0}{1}_E + \sqrt{p(1-p)} X_A \ketbra{1}{0}_E + (1-p) I_A \ketbra{1}{1}_E ).
    \end{split}
\end{equation}
Then, the classical-quantum state can be expressed as
\begin{equation}
\begin{split}
    \rho_{aE} &= \ketbra{0}{0}_a \otimes \text{Tr}_{A \boldsymbol{B}} [\ketbra{+}{+}_A \ketbra{\Psi}{\Psi}_{A\boldsymbol{B}E}] + \ketbra{1}{1}_a \otimes \text{Tr}_{A \boldsymbol{B}} [\ketbra{-}{-}_A \ketbra{\Psi}{\Psi}_{A\boldsymbol{B}E}]\\
    &= \frac{1}{2} \ketbra{0}{0}_a \otimes \rho_+ + \frac{1}{2} \ketbra{1}{1}_a \otimes \rho_-,
\end{split}
\end{equation}
where we define
\begin{equation}
    \rho_\pm \equiv p \ketbra{0}{0}_E \pm \sqrt{p (1-p)} (\ketbra{0}{1}_E + \ketbra{1}{0}_E) + (1-p) \ketbra{1}{1}_E.
\end{equation}
After Alice performs noisy preprocessing, this state becomes
\begin{equation}
\begin{split}
    \tilde{\rho}_{aE} &= \frac{1}{2} \ketbra{0}{0}_a \otimes ((1-q) \rho_+ + q \rho_-) + \frac{1}{2} \ketbra{1}{1}_a \otimes ((1-q)\rho_- + q \rho_+)\\
    &= \frac{1}{2} \ketbra{0}{0}_a \otimes \tilde{\rho}_+ + \frac{1}{2} \ketbra{1}{1}_a \otimes \tilde{\rho}_-,
\end{split}
\end{equation}
where $\tilde{\rho}_\pm = (1-q) \rho_\pm + q \rho_\mp$. The eigenvalues of $\tilde{\rho}_+$ are
\begin{equation}
    \lambda_\pm = \frac{1 \pm \sqrt{1-16 q(1-q) p (1-p)}}{2}.
\end{equation}
We find that the eigenvalues of $\tilde{\rho}_-$ are also $\lambda_\pm$. Therefore, we have
\begin{equation}
    H(E|A, x=0) = h(\lambda_+) = h(\frac{1+\sqrt{1-16p(1-p)q(1-q)}}{2}),
\end{equation}
and
\begin{equation}\label{appeq:HAE}
    H(A|E,x=0) = 1 - h(p) + h(\frac{1+\sqrt{1-16p(1-p)q(1-q)}}{2}).
\end{equation}

Next, we calculate the Bell value of the full-body Bell inequality. The reduced state shared by Alice and the Bobs is
\begin{equation}
    \rho_{A \boldsymbol{B}} = p \ketbra{D_N^{N/2}}{D_N^{N/2}} + (1-p) X_A \ketbra{D_N^{N/2}}{D_N^{N/2}} X_A.
\end{equation}
Since we have
\begin{align}
    X_A A_0 X_A = X, \quad X_A A_1 X_A = -Y,
\end{align}
the following holds
\begin{equation}
    \ev{A_0C_0} = \cos \theta_p, \quad \ev{A_0 C_1} = \cos \theta_p, \quad \ev{A_1 C_0} = (2p-1) \sin \theta_p, \quad \ev{A_1C_1} = - (2p-1) \sin \theta_p.
\end{equation}
Therefore, the Bell value can be expressed as
\begin{equation}
    S_{N, \alpha}^\text{full} = 2 (\alpha \cos \theta_p + (2p-1) \sin \theta_p).
\end{equation}
Optimizing the measurement angle $\theta_p$ gives
\begin{equation}
    S_{N, \alpha}^\text{full} = 2 \sqrt{\alpha^2 + (2p-1)^2},
\end{equation}
for $\theta_p = \arctan (\frac{2p-1}{\alpha})$. Then, the following holds 
\begin{equation}
\begin{split}
    p &= \frac{1+F_\alpha (S_{N, \alpha}^\text{full})}{2}.
\end{split}
\end{equation}
By substituting this relationship into Eq.~(\ref{appeq:HAE}), we have
\begin{equation}
    \begin{split}
        H(A|E,x=0) &= 1 - h(\frac{1+F_\alpha (S_{N, \alpha}^\text{full})}{2}) + h(\frac{1 + \sqrt{(1-2q)^2 + 4q (1-q) F_\alpha(S_{N, \alpha}^\text{full})^2}}{2}).
    \end{split}
\end{equation}
This coincides with the analytical lower bound for $\alpha \geq 1$. For $0 < \alpha < 1$, by considering a probabilistic mixture of the above eavesdropping attack and a classical eavesdropping attack which gives $H(A|E,x=0) = h(q)$ at $S_{N, \alpha}^\text{full} = 2$, we can show the tightness of the entropy bound.

\section{E. Proof of Proposition 1}
We describe the proof of Proposition 1. We aim to show that, for a fixed even $N \geq 4$ and a probability distribution $P_\epsilon \notin \mathcal{L}_N$, the following statements hold simultaneously 
\begin{equation}\label{appeq:prop}
r(P_\epsilon) > 0, \quad\mathbf{d}(P_\epsilon, \mathcal{L}_N) \leq \frac{N-1}{2}\epsilon,
\end{equation}
where $r(P_\epsilon)$ represents a DI-CKA key rate for $P_\epsilon$.

We first derive the first inequality in Eq.~(\ref{appeq:prop}). To this end, we use the full-body Bell value which we introduce in Sec. D of the Supplemental Material:
\begin{equation}
    S_{N, \alpha}^\text{full} = \alpha \ev{A_0 (C_0 + C_1)} + \ev{A_1 (C_0 - C_1)}.
\end{equation}
Let $0 < \epsilon \leq 1$ and $\alpha = 1/\epsilon$, and define
\begin{equation}
    T_{N, \epsilon}^\text{full} = \epsilon S_{N, \alpha}^\text{full} = \ev{A_0 (C_0 + C_1)} + \epsilon \ev{A_1 (C_0 - C_1)}.
\end{equation}
Since the local bound and the quantum maximum of $S_{N, \alpha}^\text{full}$ are $\max\{ 2, 2 \alpha \}$ and $2\sqrt{1+\alpha^2}$, respectively, the local bound and the quantum maximum of $T_{N, \epsilon}^\text{full}$ become 2 and $2 \sqrt{1 + \epsilon^2}$, respectively. We use the following measurements
\begin{equation}
    A_0 = X, \quad A_1 = Y, \quad B_0^i = M(\theta), \quad B_1^i = M(-\theta), \quad  B_2^i = X,
\end{equation}
where $M(\theta) = \cos \theta X + \sin \theta Y$ and $\theta = \arctan \epsilon$. Let $P_\epsilon$ denote a probability distribution obtained by performing these measurements on the balanced Dicke state. Using
\begin{equation}
    \ev{M_A(\gamma) \bigotimes_{i=1}^{N-1} M_{B_i}(\beta)}_D = \cos (\gamma - \beta),
\end{equation}
the Bell value can be expressed as
\begin{equation}
\begin{split}
    T_{N, \epsilon}^\text{full} &= 2 (\cos \theta + \epsilon \sin \theta)\\
    &= 2 \sqrt{1 + \epsilon^2}\\
    &> 2.
\end{split}
\end{equation}
That is, for all $0 < \epsilon \leq 1$, we can achieve the quantum maximum. Define
\begin{equation}
    G_\epsilon (t) = \frac{\sqrt{t^2/4-1}}{\epsilon}.
\end{equation}
Using the analytical entropy bound for the original full-body Bell inequality, we have the following entropy bound:
\begin{equation}
    H(A|E, x=0) \geq 1-h(\frac{1+G_\epsilon (T_{N, \epsilon}^\text{full})}{2}),
\end{equation}
where we set $q = 0$ for simplicity. When the quantum maximum $T_{N, \epsilon}^\text{full} = 2\sqrt{1 + \epsilon^2}$ is achieved, $G_\epsilon(T_{N, \epsilon}^\text{full}) = 1$. Therefore, the DI-CKA key rate in the ideal setting becomes
\begin{equation}
    r (P_\epsilon) = 1 - h(\frac{N-2}{4 (N-1)}) > 0,
\end{equation}
for every $0 < \epsilon \leq 1$.

Next, we derive the second inequality in Eq.~(\ref{appeq:prop}). Define the following observable where we rotate the measurement axis by $\delta$ away from $X$,
\begin{equation}
    M(\delta) = \cos \delta X + \sin \delta Y.
\end{equation}
Corresponding projectors can be expressed as
\begin{equation}
    \Pi_{\pm 1} (\delta) = \frac{I \pm M(\delta)}{2}.
\end{equation}
Define $D = \Pi_{+1} (\delta) - \Pi_{+1} (0)$. Then, since $X^2 = Y^2 = I$ and $XY + YX = 0$, we have
\begin{equation}
    \begin{split}
        D^2 
        &= \sin^2 \frac{\delta}{2} I.
    \end{split}
\end{equation}
Since $D$ is Hermitian, its eigenvalues are $\lambda_\pm = \pm \sin \frac{\delta}{2}$. The operator norm of $D$ is therefore
\begin{equation}
    ||D||_\infty = \sin \frac{|\delta|}{2},
\end{equation}
where we assume $-\pi \leq \delta \leq \pi$. We next consider the distance between two probability distributions after one party rotates their measurement axes by $\delta$. Without loss of generality, we assume that Alice rotates her measurement axis. Let $\rho_{A|\boldsymbol{b}}$ denote Alice's quantum state conditioned on the Bobs' measurement outcomes $\boldsymbol{b}$. Define
\begin{align}
    P_\text{after} (+1, \boldsymbol{b}) &= P(\boldsymbol{b}) \text{Tr} [\rho_{A|\boldsymbol{b}} \Pi_{+1} (\delta)],\\
    P_\text{before} (+1, \boldsymbol{b}) &= P(\boldsymbol{b}) \text{Tr} [\rho_{A|\boldsymbol{b}} \Pi_{+1} (0)],
\end{align}
where $P(\boldsymbol{b})$ is the probability of obtaining $\boldsymbol{b}$. Then, the following holds
\begin{equation}
    \begin{split}
        |P_\text{after} (+1, \boldsymbol{b}) - P_\text{before} (+1, \boldsymbol{b})| &= P(\boldsymbol{b}) |\text{Tr}[\rho_{A|\boldsymbol{b}} (\Pi_{+1} (\delta) - \Pi_{+1} (0))]|\\
        &= P (\boldsymbol{b}) |\text{Tr}[\rho_{A|\boldsymbol{b}} D]|\\
        &\leq P(\boldsymbol{b}) ||D||_\infty\\
        &=P(\boldsymbol{b}) \sin \frac{|\delta|}{2},
    \end{split}
\end{equation}
where we use $|\text{Tr} [\rho D]| \leq ||D||_\infty$ for any density matrix $\rho$ and Hermitian operator $D$ for the inequality. Since $\Pi_{-1} (\delta) = I - \Pi_{+1} (\delta)$, we have
\begin{equation}
    |P_\text{after}(+1, \boldsymbol{b}) - P_\text{before} (+1, \boldsymbol{b})| = |P_\text{after}(-1, \boldsymbol{b}) - P_\text{before} (-1, \boldsymbol{b})|.
\end{equation}
Consequently,
\begin{equation}
    \begin{split}
        \frac{1}{2} \sum_{a, \boldsymbol{b}} |P_\text{after} (a, \boldsymbol{b}) - P_\text{before} (a, \boldsymbol{b})| &= \sum_{\boldsymbol{b}} |P_\text{after} (+1, \boldsymbol{b}) - P_\text{before} (+1, \boldsymbol{b})|\\
        &\leq \sum_{\boldsymbol{b}}  P(\boldsymbol{b}) \sin \frac{|\delta|}{2}\\
        &= \sin \frac{|\delta|}{2}.
    \end{split}
\end{equation}
Thus, when one party rotates their measurement axis by $\delta$, the distance between two probability distributions is bounded by
\begin{equation}
    \mathbf{d}(P_\text{after}, P_\text{before}) \leq \sin \frac{|\delta|}{2}.
\end{equation}
We can straightforwardly generalize this result to the cases where multiple parties rotate their measurement axes. When $N_\text{rotate}$ parties rotate their measurement axes by $\delta$, the distance between probability distributions before and after the rotation is bounded by
\begin{equation}
    \mathbf{d}(P_\text{after}, P_\text{before}) \leq N_\text{rotate} \sin \frac{|\delta|}{2}.
\end{equation}
Finally, we construct a local-hidden-variable (LHV) model which reproduces the same probability distribution as the above configuration with $\epsilon \rightarrow 0$. Note that $B_0^i = B_1^i = B_2^i = X$ when $\epsilon = 0$. Let $P_0$ denote a probability distribution corresponding to this limit. We show $P_0 \in \mathcal{L}_N$ in the following. First, for the balanced Dicke state, we have
\begin{equation}
    X^{\otimes N} \ket{D_N^{N/2}} = \ket{D_N^{N/2}},
\end{equation}
i.e., the balanced Dicke state is an eigenstate of $X^{\otimes N}$ with eigenvalue of $+1$. Therefore, for the measurement outcomes of the legitimate parties, the following holds when all parties perform the $X$ measurements
\begin{equation}
    a \prod_{i = 1}^{N-1} b_i = +1,
\end{equation}
and consequently
\begin{equation}
    a = \prod_{i = 1}^{N-1} b_i.
\end{equation}
In addition, when all the Bobs perform the $X$ measurements on the balanced Dicke state, the remaining quantum state on Alice's system becomes $\ketbra{+}{+}$ or $\ketbra{-}{-}$. Therefore, when Alice performs the $Y$ measurement, she will obtain measurement outcomes $\pm 1$ with equal probabilities. Now, we construct an explicit LHV model which gives $P_0$. Let $P_X^B(\boldsymbol{b})$ denote a marginal probability distribution where all the Bobs perform the $X$ measurements and obtain outcomes $\boldsymbol{b} = (b_1, \ldots, b_{N-1})$. In addition, we use a variable $\xi$, satisfying $P(\xi = +1) = P(\xi = -1) = 1/2$. Then, we first sample the hidden variable $\lambda = (b_1, \ldots, b_{N-1}, \xi)$ with probability $P(\lambda) = \frac{1}{2} P_X^B(\boldsymbol{b})$. $\text{Bob}_i$ outputs $b_i$ for every input for all $i$. On the other hand, Alice outputs $\prod_{i=1}^{N-1}b_i$ when her input is $x = 0$ and outputs $\xi$ when $x = 1$. Since this LHV model produces the probability distribution $P_0$, this distribution is in the set of local correlations, i.e., $P_0 \in \mathcal{L}_N$. In conclusion, the distance between $P_\epsilon$ and $P_0$ can be expressed as
\begin{equation}
    \begin{split}
        \mathbf{d}(P_\epsilon, P_0) &\leq (N-1) \sin \frac{\theta}{2}\\
        &= (N-1) \sin \frac{\arctan \epsilon}{2}\\
        &\leq \frac{(N-1) \epsilon}{2},
    \end{split}
\end{equation}
where we use $\sin (\theta/2) \leq \theta/2$ and $\theta = \arctan \epsilon \leq \epsilon$ for the last inequality. This concludes the proof of Proposition 1. Consequently, for a fixed even $N$ and $\epsilon \rightarrow 0$, we have
\begin{equation}
    r(P_\epsilon) = 1-h(\frac{N-2}{4(N-1)}) > 0 ,\quad \mathbf{d}(P_\epsilon, \mathcal{L}_N) \rightarrow 0.
\end{equation}

\section{F. Noise analyses}
We analyze the effect of experimental imperfections, i.e., local depolarizing noise, detection inefficiency, and collective phase noise. We consider three different DI-CKA protocols: balanced-Dicke-based protocol with the full-body inequality, balanced-Dicke-based protocol with the two-body inequality, and GHZ-based protocol with the asymmetric parity-CHSH inequality~\cite{Ishihara2026Enhancing}. The same local depolarizing noise and detection efficiency are used for all parties. 

Each qubit undergoes an independent depolarizing channel, defined for a single-qubit state $\rho$ by
\begin{equation}
    D_{p_\text{dep}} (\rho) = (1-p_\text{dep}) \rho + p_\text{dep} \frac{I}{2}, \quad 0 \leq p_\text{dep} \leq 1.
\end{equation}
Consider an arbitrary traceless observable $M = n_1 X + n_2 Y + n_3 Z$, where $n_1, n_2$ and $n_3$ are coefficients. The local depolarizing noise turns this observable into
\begin{equation}
    \mathcal{D}_{p_\text{dep}}(M) = (1-p_\text{dep})M = v_\text{dep} M, \quad \mathcal{D}_{p_\text{dep}}(I) = I,
\end{equation}
where $v_\text{dep} = 1-p_\text{dep}$. This means that a $k$-body correlator is attenuated with the factor $ v_\text{dep}^k$.

We model each nondetection event by assigning an independent uniformly random binary outcome. For an ideal observable $M$, the effective POVM elements are
\begin{equation}
    E_{a|M} = \eta \frac{I + aM}{2} + (1-\eta) \frac{I}{2} = \frac{I + \eta a M}{2}.
\end{equation}
Then, the corresponding observable with detection efficiency $\eta$ becomes
\begin{equation}
    E_{+1|M} - E_{-1|M} = \eta M.
\end{equation}
Therefore, full-body and two-body correlators are attenuated with $(\eta v_\text{dep})^N$ and $(\eta v_\text{dep})^2$, respectively, after these imperfections.

Consider a phase rotation $U_\phi$. We define the collective phase noise as
\begin{equation}
  \mathcal{P} (\rho) = \int \dd \phi ~ P(\phi) U_\phi^{\otimes N} \rho (U_\phi^\dagger)^{\otimes N},
\end{equation}
where $P(\phi)$ is a Gaussian distribution with zero mean and variance $\sigma^2_\phi$. The balanced Dicke state is invariant under this noise, i.e., $U_\phi^{\otimes N} \ket{D_{N}^{N/2}} = \ket{D_N^{N/2}}$. On the other hand, the GHZ state becomes $U_\phi^{\otimes N} \ket{\text{GHZ}_N} = (\ket{0}^{\otimes N} + e^{i N \phi} \ket{1}^{\otimes N})/\sqrt{2}$ up to a global phase.

\subsection{Balanced-Dicke-based protocol with the two-body Bell inequality}
The ideal two-body correlators follow from the reduced state in Eq.~(\ref{appeq:twomode}):
\begin{equation}
    \ev{XX} = \ev{YY} = \frac{N}{2(N-1)}, \quad \ev{ZZ} = -\frac{1}{N-1}, \quad \ev{XZ} = \ev{ZX} = 0.
\end{equation}
Considering the local depolarizing noise and detection efficiency, these correlators are attenuated with the factor $(\eta v_\text{dep})^2$.

We use the following measurements
\begin{equation}
    A_0 = X, \quad A_1 = Z, \quad B_0^i = \cos \theta Z -\sin \theta X , \quad B_1^i = - \cos \theta Z - \sin \theta X , \quad B_2^i = X.
\end{equation}
Then, the two-body Bell value including the imperfections becomes
\begin{equation}
    S_{N, \alpha}^{\text{two/noise}} = (\eta v_\text{dep})^2 (\alpha N \sin \theta + 2 \cos \theta - \frac{1+\alpha}{4} (N^2-4) \sin^2 \theta)-\frac{1+\alpha}{2}(N-2)(1-(\eta v_\text{dep})^2).
\end{equation}

On the other hand, the $X$-basis quantum bit error rate between Alice and $\text{Bob}_i$ before noisy preprocessing can be expressed as
\begin{equation}
    Q_X = \frac{1- (\eta v_\text{dep})^2 \ev{X_AX_{B_i}}}{2}.
\end{equation}
After noisy preprocessing, the error rate is
\begin{equation}\label{appeq:noisyQBER}
    \begin{split}
        Q_X^q &= (1-q) Q_X + q(1-Q_X)\\
        &= \frac{1-(1-2q)(\eta v_\text{dep})^2 \ev{X_AX_{B_i}}}{2}.
    \end{split}
\end{equation}

Consequently, the DI-CKA key rate with the balanced Dicke state and the two-body Bell inequality becomes
\begin{equation}
    r_{N, \alpha}^\text{two} \geq \bar{g}_{q, \alpha} (S_{N, \alpha}^\text{two/noise}) - h(Q_X^q).
\end{equation}
We numerically optimize the measurement angle $\theta$, the coefficient $\alpha$, and the probability for noisy preprocessing $q$ to maximize the key rate.

\subsection{Balanced-Dicke-based protocol with the full-body Bell inequality}
We use the following measurements
\begin{equation}
    A_0 = X, \quad A_1 = Y, \quad B_0^i = \cos \theta X + \sin \theta Y, \quad B_1^i = \cos \theta X - \sin \theta Y, \quad B_2^i = X. 
\end{equation}
Then, the full-body Bell value including the imperfections becomes
\begin{equation}
    S_{N, \alpha}^\text{full/noise} = 2 (\eta v_\text{dep})^N (\alpha \cos \theta + \sin \theta).
\end{equation}
For $\theta = \arctan 1/\alpha$, the Bell value is maximized and, consequently,
\begin{equation}
    S_{N, \alpha}^\text{full/noise} = 2 (\eta v_\text{dep})^N \sqrt{1 + \alpha^2}.
\end{equation}
Therefore, the DI-CKA key rate with the balanced Dicke state and the full-body Bell inequality becomes
\begin{equation}
    r_{N, \alpha}^\text{full} \geq \bar{g}_{q, \alpha} (S_{N, \alpha}^\text{full/noise}) - h(Q_X^q).
\end{equation}
We numerically optimize $\alpha$ and $q$ to maximize the key rate.

\subsection{GHZ-based protocol with the asymmetric parity-CHSH inequality}
We use the following measurements
\begin{equation}
    A_0 = B_2^i = Z, \quad A_1 = B_0^2 = \cdots B_0^{N-1} = X, \quad B_0^1 = \cos \theta Z + \sin \theta X, \quad B_1^1 = \cos \theta Z - \sin \theta X.
\end{equation}
Here, each Bob except $\text{Bob}_1$ has a single measurement input for the Bell test. We use the following asymmetric parity-CHSH value~\cite{Ishihara2026Enhancing}
\begin{equation}
    S_{N, \alpha}^\text{parity} = \alpha \ev{A_0 (B_0^1 + B_1^1)} + \ev{A_1 (B_0^1 - B_1^1) \prod_{i=2}^{N-1} B_0^i},
\end{equation}
where the local bound and the quantum maximum are $\max \{  2, 2\alpha\}$ and $2 \sqrt{1+\alpha^2}$, respectively. For the above ideal measurements, the Bell value becomes
\begin{equation}
    S_{N, \alpha}^\text{parity} = 2 \alpha \cos \theta \ev{Z_AZ_{B_1}} + 2\sin \theta \ev{X_A \prod_{i=1}^{N-1} X_{B_i}}.
\end{equation}

In the ideal scenario, the relevant two correlators satisfy
\begin{equation}
    \ev{Z_A Z_{B_1}} =  \ev{X_A \prod_{i = 1}^{N-1} X_{B_i}} = 1.
\end{equation}
Since the phase rotation $U_\phi$ does not affect the $Z$ measurement, the two-body correlator with the imperfections becomes
\begin{equation}
    \ev{Z_A Z_{B_1}}^\text{noise} = (\eta v_\text{dep})^2.
\end{equation}
On the other hand, a density matrix corresponding to the GHZ state after the collective phase rotation can be expressed as
\begin{equation}
    \rho_\phi = \frac{1}{2} (\ketbra{\mathbf{0}}{\mathbf{0}} + \ketbra{\mathbf{1}}{\mathbf{1}} + e^{-iN\phi}\ketbra{\mathbf{0}}{\mathbf{1}} + e^{iN\phi}\ketbra{\mathbf{1}}{\mathbf{0}}),
\end{equation}
where $\ket{\mathbf{0}} = \ket{0}^{\otimes N}$ and $\ket{\mathbf{1}} = \ket{1}^{\otimes N}$. Averaging over $\phi$ gives
\begin{equation}
\begin{split}
    \bar{\rho} &= \int \dd \phi ~ P(\phi)\rho_\phi\\
    &= \int \dd \phi ~ \frac{1}{\sqrt{2 \pi \sigma^2_\phi} } \exp(- \frac{\phi^2}{2 \sigma^2_\phi}) \rho_\phi  \\
    &= \frac{1}{2} (\ketbra{\mathbf{0}}{\mathbf{0}} + \ketbra{\mathbf{1}}{\mathbf{1}} + \chi_N\ketbra{\mathbf{0}}{\mathbf{1}} + \chi_N\ketbra{\mathbf{1}}{\mathbf{0}}),
\end{split}
\end{equation}
where
\begin{equation}
    \begin{split}
        \chi_N &= \int \dd \phi ~ P(\phi) e^{iN\phi}\\
        &= \exp(-\frac{N^2 \sigma^2_\phi}{2}).
    \end{split}
\end{equation}
Therefore, the full-body correlator with the imperfections becomes
\begin{equation}
\ev{X_A \prod_{i = 1}^{N-1} X_{B_i}}^\text{noise} = \chi_N (\eta v_\text{dep})^N.
\end{equation}
Consequently, the asymmetric parity-CHSH value becomes
\begin{equation}
    S_{N, \alpha}^\text{parity/noise} = 2 (\alpha (\eta v_\text{dep})^2 \cos \theta + \chi_N (\eta v_\text{dep})^N \sin \theta).
\end{equation}
Optimizing the measurement angle $\theta$ gives
\begin{equation}
    S_{N, \alpha}^\text{parity/noise} = 2 \sqrt{\alpha^2 (\eta v_\text{dep})^4 + \chi_N^2 (\eta v_\text{dep})^{2N}}.
\end{equation}

Since $\ev{Z_AZ_{B_i}} = (\eta v_\text{dep})^2$, the $Z$-basis quantum bit error rate before noisy preprocessing is
\begin{equation}
    Q_Z = \frac{1- \ev{ZZ}}{2} = \frac{1-(\eta v_\text{dep})^2}{2}.
\end{equation}
After noisy preprocessing, this error rate is
\begin{equation}
    \begin{split}
        Q_Z^q &= (1-q) Q_Z + q (1-Q_Z)\\
        &= \frac{1-(1-2q)(\eta v_\text{dep})^2}{2}.
    \end{split}
\end{equation}
Note that the quantum bit error rate becomes zero in the ideal setting, i.e., $\eta = v_\text{dep} =  1$ for $q = 0$.

By using the result in Ref.~\cite{Ishihara2026Enhancing}, the key rate of the GHZ-based protocol can be expressed as
\begin{equation}
    r_{N, \alpha}^\text{GHZ} \geq \bar{g}_{q, \alpha} (S_{N, \alpha}^\text{parity/noise}) - h(Q_Z^q).
\end{equation}
We numerically optimize the coefficient $\alpha$ and the probability for noisy preprocessing $q$ to maximize the key rate.

\begin{figure}[tbp]
\includegraphics[keepaspectratio, scale=0.55]{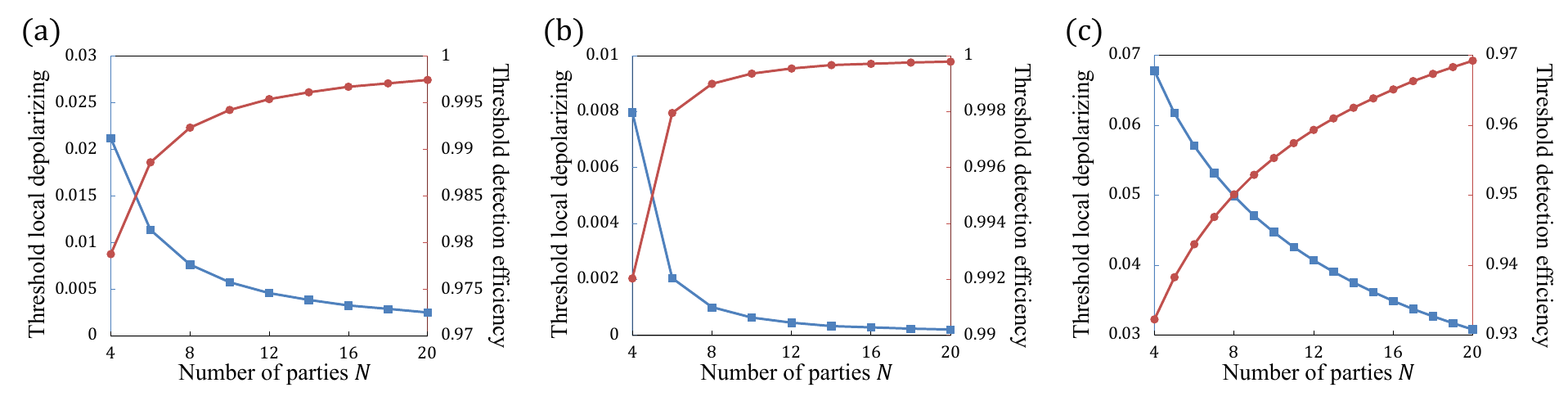}
\caption{\label{fig:NoiseAll}
Noise thresholds versus the number of parties $N$: (a) the full-body balanced-Dicke-based protocol, (b) the two-body balanced-Dicke-based protocol, and (c) the GHZ-based protocol with the asymmetric parity-CHSH inequality. The blue squares, read against the left axis, show the maximum tolerable local depolarizing noise $p_\text{dep}$; the red circles, read against the right axis, show the minimum required detection efficiency $\eta$.}
\end{figure}

\subsection{Numerical results}
We numerically evaluate the optimized key rates and determine the noise thresholds. The local depolarizing threshold is the largest value of $p_\text{dep}$ compatible with a positive key rate, whereas the detection efficiency threshold is the minimum required efficiency $\eta$ for a positive key rate. Figure~\ref{fig:NoiseAll} compares these thresholds for the three DI-CKA protocols. The GHZ-based protocol has the highest tolerance to local depolarizing noise and detection inefficiency, followed by the full-body Dicke-based protocol and then the two-body Dicke-based protocol. For the Dicke-based protocols, the nonzero error-correction cost in the ideal setting reduces the tolerable noise margin. The two-body protocol has an additional limitation: high conditional von Neumann entropy is certified at small $\alpha$, whereas the separation between the local and quantum bounds of the inequality becomes smaller as $\alpha$ decreases. Consequently, the protocol becomes more sensitive to local depolarizing noise and detection loss.

\end{document}